\documentclass[final,3p,times,twocolumn,12pt]{elsarticle}

\usepackage{amssymb}

\biboptions{sort&compress}

\usepackage{amsmath}
\usepackage[hidelinks]{hyperref}
\usepackage{color}
\usepackage{siunitx}

\journal{JQSRT}

\begin{document}

\begin{frontmatter}



\title{Thermal radiative energy exchange between a closely-spaced linear chain of spheres and its environment}


\author{Braden Czapla}
\author{Arvind Narayanaswamy}
\ead{arvind.narayanaswamy@columbia.edu}

\address{Department of Mechanical Engineering, Columbia University, New York, NY 10027, USA}

\begin{abstract}
In this work, we present expressions for radiative heat transfer between pairs of spheres in a linear chain and between individual spheres and their environment. The expressions are valid for coated spheres of arbitrary size, spacing, and isotropic optical properties. The spheres may be small and closely-spaced, which violates the assumptions foundational to classical radiative transfer. We validate our results against existing formulations of radiative heat transfer, namely the thermal discrete dipole and boundary element methods. Our results have important implications for the modeling and interpretation of near-field radiative heat transfer experiments between spherical bodies.
\end{abstract}


\end{frontmatter}



\section{Introduction} \label{sec:Introduction}
In some sense, the theory of radiative heat transfer is complete: use Maxwell's equations to determine the Poynting vector and, ultimately, the flow of electromagnetic energy between bodies. For objects and separation distances much larger than the dominant thermal wavelength ($\lambda_{T} \approx$ 10 \si{\micro\meter} at room temperature), radiative heat transfer between engineering surfaces can be simplified to problems involving Planck's blackbody distribution (or the Stefan-Boltzmann law), emissivities, and view factors \cite{Howell2011}. These classical concepts carry readily apparent physical significance.

The story changes when objects and separation distances become comparable to the dominant thermal wavelength and ``near-field effects" such as the diffraction, interference, and tunneling of electromagnetic waves create pronounced deviations from what classical theory predicts. A number of methods to solve Maxwell's equations and account for those effects in near-field radiative heat transfer (NFRHT) have been developed. The most common of these methods involve dyadic Green's functions (DGFs) \cite{Joulain2005, Volokitin2007, Francoeur2008, Narayanaswamy2013a}, spectral densities \cite{Kruger2012}, fluctuating surface currents/boundary element methods (BEM) \cite{Rodriguez2012}, or thermal discrete dipole approximations (T-DDA) \cite{Edalatpour2014}, each with their own advantages and limitations. But even with these formulations of the problem, physical insight can be lacking.

To that end, a great deal of research has been directed towards determining explicit formulas for heat transfer in experimentally important thermal systems. Certain problems are best attacked using DGF and spectral density methods, such as NFRHT in plane-plane \cite{Polder1971}, sphere-sphere \cite{Narayanaswamy2008, Mackowski2008, Kruger2012, Czapla2017}, and sphere-plane \cite{Otey2011, Kruger2012} configurations, where the electromagnetic fields are easily described by vector wave eigenfunction expansions. Thermal systems which do not yield to those methods often require methods that involve surface or volume meshing, such as BEM and T-DDA. Those methods have allowed investigations of cone-plane \cite{Rodriguez2013, Edalatpour2016} and cube-cube \cite{Edalatpour2014} configurations, to name a few.

The goal of this work is simple: provide explicit formulas which completely describe a thermal system consisting of spheres in a linear chain. While this system has been previously investigated for light scattering \cite{Fuller1988, Li2003, Wei2004, Chen2006, Lee2013} and radiative transfer in the dipole limit \cite{Ben-Abdallah2011, Dong2017a, Kathmann2018}, it has not yet been characterized for NFRHT between spheres of arbitrary size, with or without spherical layers. The main results of this work are numerically exact expressions for NFRHT between pairs of spheres in a linear chain, and between any sphere in a chain and its environment. Furthermore, motivated by recent experimental measurements of the Casimir force between microspheres \cite{Garrett2018} and informed by our new expressions, we developed a thermal model for a potential NFRHT experiment between two spheres and provide suggestions on how to best interpret the measured sphere-sphere heat transfer.

The structure of the paper is as follows: first, in Sec. \ref{Geometry}, the geometry of the linear chair is described and geometrical variables are defined. Next, in Sec. \ref{sec:Theory}, the theoretical treatment is discussed and the DGFs of the system are used to determine a formula for energy exchange between spheres in a chain and their environment. Then, in Sec. \ref{sec:CrossValidationofTheoreticalResults}, the formulas are validated against existing results in the literature. Finally, in Sec. \ref{sec:Implications}, the implications of the present work are discussed as they relate to NFRHT experiments between spherical objects.

\section{Geometry} \label{Geometry}
\begin{figure}
\includegraphics[width=3 in]{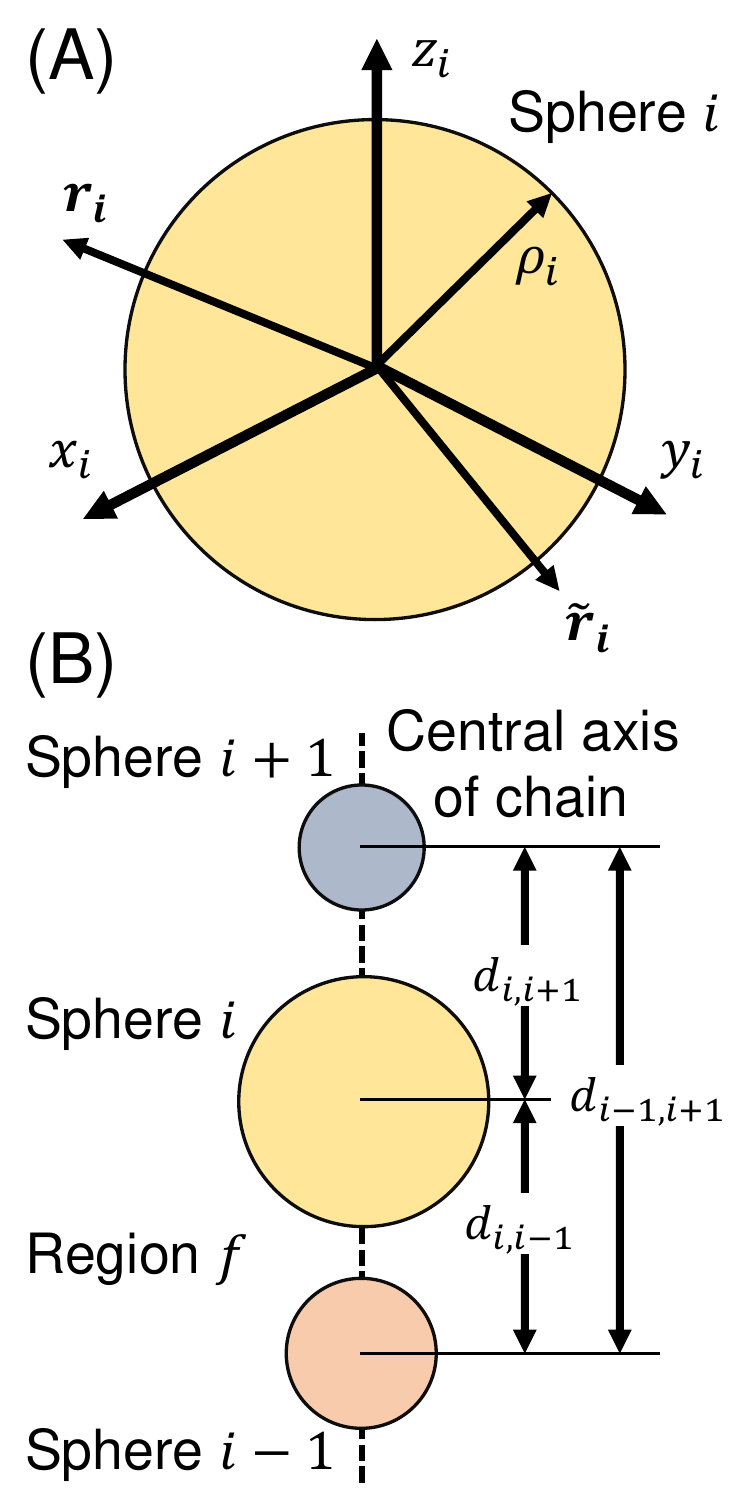}
\caption{\label{fig:Geometry} Configuration of spheres in chain. (A) Single sphere $i$ and its associated coordinate system. (B) Section from a linear chain of spheres embedded in region $f$.}
\end{figure}

The configuration of the spheres is shown in Fig. \ref{fig:Geometry}. In Fig. \ref{fig:Geometry}A, an individual sphere, labeled sphere $i$, is depicted. Sphere $i$ has an outer radius, $\rho_{i}$. Internally, it may be homogeneous or composed of spherically symmetric layers. Coordinate system $i$ is fixed to the center of sphere $i$. Any position vector, $\boldsymbol{r}$ or $\widetilde{\boldsymbol{r}}$, when written in coordinate system $i$, is denoted $\boldsymbol{r}_{i}$ or $\widetilde{\boldsymbol{r}}_{i}$, respectively. 

Figure \ref{fig:Geometry}B depicts a section of a linear chain of $N_{s}$ spheres, which is embedded in free space (referred to as region $f$). The $z$-axis of coordinate system $i$ ($1 \le i \le N_{s}$) is aligned down the central axis of the chain. The $x$-axes of all coordinate systems are parallel, and similarly so are the $y$-axes.

The spheres are numbered $1$ through $N_{s}$, such that their labels increase along the positive $z$-direction (of any coordinate system). Any two spheres, $i$ and $j$, are separated by center-to-center distance $d_{i,j}$ and the minimum separation between them is $D_{i,j} = | d_{i,j} | - \rho_{i} - \rho_{j}$ (not depicted explicitly in Fig. \ref{fig:Geometry}). The sign of $d_{i,j}$ is positive if $j > i$ and negative if $j < i$.

\section{Theory} \label{sec:Theory}
The net radiative transfer between any two objects may be determined if the DGFs, which describe the electromagnetic fields created by any distribution of sources, are known. To solve electromagnetism problems, the electric and magnetic DGFs ($\overline{\overline{\boldsymbol{G}}}_{e}(\boldsymbol{r}; \widetilde{\boldsymbol{r}})$ and $\overline{\overline{\boldsymbol{G}}}_{m}(\boldsymbol{r}; \widetilde{\boldsymbol{r}})$, respectively) are required. Both variants of DGFs take two position vector arguments. The significance and locations of the position vectors will be explained later in this section.

For a given angular frequency, $\omega$, we may define a transmissivity function for energy transfer from object $j$ to object $i$, $\tau_{j \rightarrow i}(\omega)$, based solely on the DGFs. The mathematical definition of $\tau_{j \rightarrow i}(\omega)$ is given as Eq. 30 by Narayanaswamy and Zheng \cite{Narayanaswamy2013a} for isothermal objects with isotropic optical properties. In lieu of providing the lengthy definition, key features of the transmissivity function will be described. The transmissivity function is a function of only optical and geometric parameters, which are encoded into the DGFs. Though the exchange of thermal energy is a volumetric phenomenon, the properties of Maxwell's equations allow us to write the formulas for heat transfer in terms of surface integrals. The transmissivity function contains two surface integrals, over the surfaces of the two objects between which heat transfer is being investigated. To evaluate these integrals easily, the two position vector arguments of the DGFs should be located on the surfaces of interest, one on each surface. A surface integral formulation of NFRHT has the advantage of reducing the computations required, a property which is exploited by the BEM \cite{Rodriguez2012}.

In some ways, the double surface integral formula for the transmissivity function appears similar to the formula for a view factor \cite{Howell2011} in classical radiative transfer. The view factor, however, is a purely geometric property of just two objects, irrespective of any other objects present, whereas the DGFs contained within the transmissivity function automatically account for the effects of all objects present. Furthermore classical approaches such as thermal circuits or Gebhart factors \cite{Gebhart1961}, when using view factors in their calculations, assume objects emit diffusely with well-defined emissivities. DGF approaches, and therefore the transmissivity function, are exact in all situations where Maxwell's equations are valid, which includes objects with super-Planckian (greater than that of a blackbody) effective emissivities.

The transmissivity function is defined such that the spectral conductance between objects $j$ and $i$, $G_{j \rightarrow i}$, is given by
\begin{equation} \label{eq:SpectralConductance}
\begin{split}
G_{j \rightarrow i}(\omega, T) &= \lim_{T_{i},T_{j} \rightarrow T} \frac{Q_{j \rightarrow i}(\omega)}{T_{j} - T_{i}}
\\
&= k_{b} \left( \frac{\mathcal{X}}{\sinh{\mathcal{X}}} \right)^{2} \tau_{j \rightarrow i}(\omega) 
\end{split}
\end{equation}
where $\mathcal{X}=\hbar \omega / 2 k_{b} T$, $\hbar$ is the reduced Plank constant, $k_{b}$ is the Boltzmann constant, $T$ is the thermodynamic temperature, and $Q_{j \rightarrow i}(\omega) d\omega$ is the contribution to the total heat transfer from radiation at angular frequencies between $\omega$ and $\omega + d\omega$.

The total conductance between objects $j$ and $i$, $G_{t,j \rightarrow i}$, is defined as
\begin{equation} \label{eq:TotalConductance}
G_{t,j \rightarrow i}(T) = \int_{0}^{\infty} \frac{d\omega}{2\pi} G_{j \rightarrow i}(\omega, T) \\
\end{equation}

Spectral conductance is often expressed in different units in different papers, but changes of variable can be performed to transform between units. A definition of spectral conductance is valid so long as the integral definition of total conductance computes to the same value, regardless of the units used. For example, in Fig. \ref{fig:ThreeSpheres_SCUFFEM} we present $\tau_{j \rightarrow i}(\lambda)$ in units of \si{\per\second \per\micro\meter}, where $\lambda = 2 \pi c / \omega$ is the vacuum wavelength and $c$ is speed of light in vacuum. This version of spectral conductance is obtained from $\tau_{j \rightarrow i}(\lambda) = c \lambda^{-2} \tau_{j \rightarrow i}(\omega)$ and the corresponding integral definition for total conductance is
\begin{equation} \label{eq:TotalConductance_wavelength}
\begin{split}
G_{t,j \rightarrow i}(T) &= \int_{0}^{\infty} G_{j \rightarrow i}(\lambda, T) d\lambda
\\
&= \int_{0}^{\infty} k_{b} \left( \frac{\mathcal{X}}{\sinh{\mathcal{X}}} \right)^{2} \tau_{j \rightarrow i}(\lambda) d\lambda 
\end{split}
\end{equation}

As explained by Narayanaswamy and Zheng \cite{Narayanaswamy2013a}, computing the transmissivity function is actually an exercise in determining DGFs. The DGFs for just two coated spheres are discussed in depth by Czapla and Narayanaswamy \cite{Czapla2017}. In this work, we have expanded on that methodology to determine DGFs for any number of non-intersecting spheres. The new DGFs reduce to those of Czapla and Narayanaswamy \cite{Czapla2017} in the case of two spheres, but also allow access to longer chains to probe the effect of additional spheres on NFRHT. The appropriate electric DGF is given by
\begin{equation}\label{eq:ElectricDGF}
\begin{split}
& \overline{\overline{\boldsymbol{G}}}_{e}(\boldsymbol{r}; \boldsymbol{\widetilde{r}})
= ik_{f} \sum\limits_{m=-\infty}^{\infty} \sum\limits_{l=\widetilde{m}}^{\infty} (-1)^{m}
\\*
& \times \left\{ \! \! \begin{array}{l}
\boldsymbol{M}_{l m}^{(3)}(k_{f} \boldsymbol{r_{i}})
\left[ \! \! \begin{array}{l} \boldsymbol{M}_{l, -m}^{(1)}(k_{f} \boldsymbol{\widetilde{r}_{j}}) \\ + R_{l}^{(M)}(\rho_{j}) \boldsymbol{M}_{l, -m}^{(3)}(k_{f} \boldsymbol{\widetilde{r}_{j}}) \end{array} \! \! \right]
\\
+ \boldsymbol{N}_{l m}^{(3)}(k_{f} \boldsymbol{r_{i}})
\left[ \! \! \begin{array}{l} \boldsymbol{N}_{l, -m}^{(1)}(k_{f} \boldsymbol{\widetilde{r}_{j}}) \\ + R_{l}^{(N)}(\rho_{j}) \boldsymbol{N}_{l, -m}^{(3)}(k_{f} \boldsymbol{\widetilde{r}_{j}}) \end{array} \! \! \right]
\end{array} \! \! \right\}
\\
& + ik_{f} \sum\limits_{p=1}^{N_{s}} \sum\limits_{m=-\infty}^{\infty} \sum\limits_{l=\widetilde{m}}^{\infty} \sum_{\nu = \widetilde{m}}^{\infty} (-1)^{m}
\\*
& \times \left\{ \! \! \begin{array}{l} \left[ \! \! \begin{array}{l}
R_{\nu}^{(M)}(\rho_{p}) V_{l, \nu,m}^{M,M,p,j} \boldsymbol{M}_{\nu m}^{(3)}(k_{f} \boldsymbol{r_{p}})
\\[8 pt]
+ R_{\nu}^{(N)}(\rho_{p}) V_{l, \nu,m}^{N,M,p,j} \boldsymbol{N}_{\nu m}^{(3)}(k_{f} \boldsymbol{r_{p}})
\end{array} \! \! \right] \\
\otimes \left[ \! \! \begin{array}{l} \boldsymbol{M}_{l, -m}^{(1)}(k_{f} \boldsymbol{\widetilde{r}_{j}}) \\ + R_{l}^{(M)}(\rho_{j}) \boldsymbol{M}_{l, -m}^{(3)}(k_{f} \boldsymbol{\widetilde{r}_{j}}) \end{array} \! \! \right]
\\[20 pt]
+ \left[ \! \! \begin{array}{l}
R_{\nu}^{(M)}(\rho_{p}) V_{l, \nu,m}^{M,N,p,j} \boldsymbol{M}_{\nu m}^{(3)}(k_{f} \boldsymbol{r_{p}})
\\[8 pt]
+ R_{\nu}^{(N)}(\rho_{p}) V_{l, \nu,m}^{N,N,p,j} \boldsymbol{N}_{\nu m}^{(3)}(k_{f} \boldsymbol{r_{p}})
\end{array} \! \! \right] \\
\otimes \left[ \! \! \begin{array}{l} \boldsymbol{N}_{l, -m}^{(1)}(k_{f} \boldsymbol{\widetilde{r}_{j}}) \\ + R_{l}^{(N)}(\rho_{j}) \boldsymbol{N}_{l, -m}^{(3)}(k_{f} \boldsymbol{\widetilde{r}_{j}}) \end{array} \! \! \right]
\end{array} \! \! \right\}
\end{split}
\end{equation}
where $k=(\omega/c) \sqrt{\varepsilon \mu}$; $\varepsilon$ and $\mu$ are the relative permittivity and permeability, respectively; $\times$ denotes a line break (not a cross product); $\otimes$ denotes a dyadic product \cite{Lai2009}; $m$, $l$, $\nu$, and $p$ are summation indices; and $\widetilde{m}=\max{\left\{|m|,1\right\}}$.

$\boldsymbol{M}_{l m}^{(n)}(k \boldsymbol{r})$ and $\boldsymbol{N}_{l m}^{(n)}(k \boldsymbol{r})$ are vector spherical waves (VSWs) of order $(l,m)$ and argument $k \boldsymbol{r}$. VSWs represent incoming (outgoing) waves for $n=1$ ($n=3$). $R_{l}^{(M)}(\rho_{i})$ and $R_{l}^{(N)}(\rho_{i})$ are the effective Mie reflection coefficients at the interface between region $f$ and sphere $i$ for $\boldsymbol{M}_{l m}^{(1)}(k_{f} \boldsymbol{r_{i}})$ and $\boldsymbol{N}_{l m}^{(1)}(k_{f} \boldsymbol{r_{i}})$ waves, respectively. See Czapla and Narayanaswamy \cite{Czapla2017} for the definitions of VSWs and Mie reflection coefficients, which are exactly identical to those used in this work.

$V_{l, \nu, m}^{X,Y,i,j}$ are scattered field coefficients. $X$ and $Y$ may take values of $M$ or $N$, and are determined by the left and right VSWs in a dyadic product, respectively. $i$ and $j$ may take any value from 1 to $N_{s}$, and are determined by the coordinate systems used to express the position vectors arguments of the  left and right VSWs in a dyadic product, respectively. For example, $V_{l, \nu, m}^{X,Y,i,j} \boldsymbol{X}_{\nu m}^{(3)}(k_{f} \boldsymbol{r_{i}}) \boldsymbol{Y}_{l, -m}^{(1)}(k_{f} \boldsymbol{\widetilde{r}_{j}})$ is a valid combination of coefficient and dyadic product.

For given values of $m$ and $j$, $V_{l, \nu, m}^{X,Y,i,j}$ may be obtained by solving the coupled set of linear equations generated from all possible combinations of $X=M$ or $N$, $Y = M$ or $N$, and $i = 1,2,...,$ or $N_{s}$ using
\begin{equation}\label{eq:ScatteredFieldEquations}
\begin{split}
&  \sum_{p=1}^{N_{s}} \sum_{n = \widetilde{m}}^{\infty}
\left[ \begin{array}{r}
V_{l,n,m}^{M,Y,p,j} R_{n}^{(M)}(\rho_{p}) C_{n,\nu,m}^{X,M,i,p}
\\
+ V_{l,n,m}^{N,Y,p,j} R_{n}^{(N)}(\rho_{p}) C_{n,\nu,m}^{X,N,i,p}
\end{array} \right]
\\
& \enskip = V_{l, \nu, m}^{X,Y,i,j} - C_{l, \nu, m}^{X,Y,i,j}
\end{split}
\end{equation}
where
\begin{equation}\label{eq:TranslationCoefficients}
C_{l, \nu, m}^{X,Y,i,j} \! = \! \left\{ \! \!
\begin{array}{cll}
0 & \! \! \text{if} & \! \! i \! = \! j
\\
A_{\nu,m}^{l,m}\left( k_{f} d_{j,i} \right) & \! \! \text{if} & \! \! i \! \neq \! j \text{ and } X \! = \! Y
\\
\! B_{\nu,m}^{l,m}\left( k_{f} d_{j,i} \right) & \! \! \text{if} & \! \! i \! \neq \! j \text{ and } X \! \neq \! Y
\end{array}
\! \! \right.
\end{equation}
and $A_{\nu,m}^{l,m}$ and $B_{\nu,m}^{l,m}$ are vector addition translation coefficients \cite{Chew1992, Chew1995, Kim2004a, Dufva2008}. The linear system of equations for scattered field coefficients is obtained by evaluating boundary conditions between the surfaces of the spheres and region $f$. It contains information on the optical properties and internal configurations of the spheres (encoded by the Mie reflection coefficients) and the geometric configuration of the ensemble of spheres (encoded by the vector additional translation coefficients). Further detail on how to solve this linear system is provided in \ref{ap:SolutionToLinearSystem}.

At this point, it is important to point out that the electric DGF given in Eq. \ref{eq:ElectricDGF} is valid for any cluster of spheres, not just a linear chain. Equations \ref{eq:ScatteredFieldEquations} and \ref{eq:TranslationCoefficients}, however, require modification before they may be applied to arbitrary clusters. See Mackowski and Mishchenko \cite{Mackowski1996} for further details on light scattering in clusters of spheres.

The magnetic DGF may be obtained from Eq. \ref{eq:ElectricDGF} by substituting $M \leftrightarrow N$ in every superscript of the reflection and VSW coefficients. Additionally, we define $\overline{\overline{\boldsymbol{G}}}_{E} = \nabla \times \overline{\overline{\boldsymbol{G}}}_{e}$ and $\overline{\overline{\boldsymbol{G}}}_{M} = \nabla \times \overline{\overline{\boldsymbol{G}}}_{m}$, where the curl operates on the first term only in each dyadic product summand of the DGFs. $\overline{\overline{\boldsymbol{G}}}_{e}$, $\overline{\overline{\boldsymbol{G}}}_{m}$, $\overline{\overline{\boldsymbol{G}}}_{E}$, and $\overline{\overline{\boldsymbol{G}}}_{M}$ are the four DGFs required to evaluate the transmissivity function.

As stated earlier, the position vector arguments of the DGFs are each located on one of the two surfaces over which the surface integrals in the transmissivity function are computed. For ease of computation, it is important to express those position vectors in the coordinate system most convenient to that goal. The choice of coordinate system therefore varies, depending on the objects between which the transmissivity function is being computed.

To obtain heat transferred from sphere $j$ to sphere $i$, $\widetilde{\boldsymbol{r}}$ must be written in the $j$-coordinate system and be located just outside the surface of sphere $j$. Similarly, $\boldsymbol{r}$ must be written in the $i$-coordinate system and be located just outside the surface of sphere $i$. Hence, we choose to represent the DGFs as $\overline{\overline{\boldsymbol{G}}}(\boldsymbol{r}_{i}; \widetilde{\boldsymbol{r}}_{j})$. The DGFs may then be simplified by using Eq. \ref{eq:ScatteredFieldEquations} to remove explicit appearances of the vector addition translation coefficients.

To obtain heat transferred from a sphere $j$ to its environment, $\widetilde{\boldsymbol{r}}$ must remain on the surface of sphere $j$ and $\boldsymbol{r}$ must lie on a large fictitious surface, whose size expands to infinity. For ease of computation, we choose the fictitious surface to be spherical. For any value of $i$, VSWs with arguments of $k_{f} \boldsymbol{r}_{i}$ asymptotically become equal as $| \boldsymbol{r} |/d_{1,N_{s}} \rightarrow \infty$. For this reason, $\boldsymbol{r}$ may be written in the coordinate system of any sphere. For ease, we will also write $\boldsymbol{r}$ in the $j$-coordinate system. Hence, we choose to represent the DGFs as $\overline{\overline{\boldsymbol{G}}}(\boldsymbol{r}_{j}; \widetilde{\boldsymbol{r}}_{j})$.

Using the simplified DGFs and assuming region $f$ is non-dissipative, the transmissivity function from sphere $j$ to sphere $i$ is given by 
\begin{equation} \label{eq:Transmissivity_SphereSphere}
\begin{split}
& \tau_{j \rightarrow i}\left( \omega \right)
= \left( k_{f} \rho_{i} \right)^{2} \left( k_{f} \rho_{j} \right)^{2} \sum\limits_{m=-\infty}^{\infty} \sum\limits_{l=\widetilde{m}}^{\infty}\sum_{\nu = \widetilde{m}}^{\infty}
\\*
& \times \left[ \! \! \begin{array}{r}
\left[ \! \! \begin{array}{r}
\epsilon_{\nu}^{(M)}(\rho_{i})
\left| V_{l,\nu,m}^{M,M,i,j} \right|^{2}
\\
+ \epsilon_{\nu}^{(N)}(\rho_{i})
\left| V_{l,\nu,m}^{N,M,i,j} \right|^{2}
\end{array} \! \! \right]
\epsilon_{l}^{(M)}(\rho_{j})
\\
+ \left[ \! \! \begin{array}{r}
\epsilon_{\nu}^{(M)}(\rho_{i})
\left| V_{l,\nu,m}^{M,N,i,j} \right|^{2}
\\
+ \epsilon_{\nu}^{(N)}(\rho_{i})
\left| V_{l,\nu,m}^{N,N,i,j} \right|^{2}
\end{array} \! \! \right]
\epsilon_{l}^{(N)}(\rho_{j})
\end{array} \! \! \right]
\end{split}
\end{equation}
where
\begin{equation}
\epsilon_{\nu}^{(P)}(\rho_{i}) = \frac{2}{\left( k_{f} \rho_{i} \right)^{2}} \left[ \Re \left( R_{\nu}^{(P)}(\rho_{i}) \right) + \left| R_{\nu}^{(P)}(\rho_{i}) \right|^{2} \right]
\end{equation}
and $P=M$ or $N$. $\Re(z)$ and $\left| z \right|$ denote the real part and magnitude of complex number $z$, respectively. This result was reported previously by Czapla and Narayanaswamy \cite{Czapla2017} for the two sphere case but here we show that the same formula, with modified scattered field coefficients, holds true for any number of spheres in a chain.

$\epsilon_{\nu}^{(P)}(\rho_{i})$ is defined such that the spectral emissivity of an isolated sphere \cite{Kattawar1970} is given by
\begin{equation}\label{eq:SpectralEmissivity}
\epsilon\left( \omega \right)
= \sum\limits_{m=-\infty}^{\infty} \sum\limits_{l=\widetilde{m}}^{\infty}
\left[ \epsilon_{l}^{(M)}(\rho_{i}) + \epsilon_{l}^{(N)}(\rho_{i}) \right]
\end{equation}
The spectral emissivity of an isolated sphere determined by Kattawar and Eisner \cite{Kattawar1970} can alternatively be derived using the DGF formalism from this work. To do so, we first compute the transmissivity function from a single sphere to a large ficticious spherical surface surrounding it. For $N_{s}=1$, Eq. \ref{eq:ScatteredFieldEquations} tells us $V_{l, n, m}^{X,Y,1,1}=0$ for all values of $l$, $n$, $m$, $X$, and $Y$. The transmissivity function for an isolated sphere $j$ to its environment, $\tau_{j \rightarrow E}^{iso}$, is given by
\begin{equation}\label{eq:Transmissivity_IsoSphereEnv}
\begin{split}
& \tau_{j \rightarrow E}^{iso} (\omega) = 2 \left( k_{f} \rho_{j} \right)^{2} \sum\limits_{m=-\infty}^{\infty} \sum\limits_{l=\widetilde{m}}^{\infty}
\\
& \times \left[ \epsilon_{l}^{(M)}(\rho_{j}) + \epsilon_{l}^{(N)}(\rho_{j}) \right]
\end{split}
\end{equation}

Next, we determine the transmissivity function for a blackbody sphere, $\tau_{j \rightarrow E, BB}^{iso}$, given by \cite{Narayanaswamy2013a} 
\begin{equation}\label{eq:Transmissivity_BBSphereEnv}
\begin{split}
\tau_{j \rightarrow E, BB}^{iso} (\omega) = \frac{k_{f}^{2}}{2 \pi} A_{j} F_{j \rightarrow E}
= 2 \left( k_{f} \rho_{j} \right)^{2}
\end{split}
\end{equation}
where $A_{j} = 4 \pi \rho_{j}^{2}$ is the surface area of sphere $j$ and $F_{j \rightarrow E}$ is the view factor from sphere $j$ to its environment ($F_{j \rightarrow E}=1$ for an isolated sphere in a large enclosure). Dividing the true amount of heat emitted (Eq. \ref{eq:Transmissivity_IsoSphereEnv}) by the amount emitted by a blackbody (Eq. \ref{eq:Transmissivity_BBSphereEnv}) yields the emissivity (Eq. \ref{eq:SpectralEmissivity}).

In the presence of additional spheres, the conductance from sphere $j$ to its environment is given by 
\begin{equation}\label{eq:Transmissivity_SphereEnv}
\begin{split}
& \tau_{j \rightarrow E}(\omega) = \tau_{j \rightarrow E}^{iso} (\omega)
\\
& +
4 \left( k_{f} \rho_{j} \right)^{2} \sum\limits_{m=-\infty}^{\infty} \sum\limits_{l=\widetilde{m}}^{\infty}
\\
& \times \left\{ \begin{array}{r}
\Re \left[ S_{l,l,m}^{M,M,p,j} \right] \epsilon_{l}^{(M)}(\rho_{j})
\\
+ \Re \left[ S_{l,l,m}^{N,N,p,j} \right] \epsilon_{l}^{(N)}(\rho_{j})
\end{array} \right\}
\\
& +
2 \left( k_{f} \rho_{j} \right)^{2} \sum\limits_{m=-\infty}^{\infty} \sum\limits_{l=\widetilde{m}}^{\infty} \sum_{\nu = \widetilde{m}}^{\infty}
\\
& \times \left\{ \begin{array}{r}
\left[ \begin{array}{r}
\left| S_{l,\nu,m}^{M,M,p,j} \right|^{2}
\\
+ \left| S_{l,\nu,m}^{N,M,p,j} \right|^{2}
\end{array} \right] \epsilon_{l}^{(M)}(\rho_{j})
\\
+ \left[ \begin{array}{r}
\left| S_{l,\nu,m}^{M,N,p,j} \right|^{2}
\\
+ \left| S_{l,\nu,m}^{N,N,p,j} \right|^{2}
\end{array} \right] \epsilon_{l}^{(N)}(\rho_{j})
\end{array} \right\}
\end{split}
\end{equation}
where $S_{l,\nu,m}^{X,Y,i,j} = \sum_{i=1}^{N_{s}} R_{\nu}^{(X)}(\rho_{i}) V_{l,\nu,m}^{X,Y,i,j}$.

Equations \ref{eq:Transmissivity_SphereSphere} and \ref{eq:Transmissivity_SphereEnv} are the main results of this work. It is important to note that these equations are valid, in principle, for spheres of any outer radii, number of coatings, separation gaps, and isotropic optical properties. This stands in contrast to the many prior works where explicit results for sphere-sphere NFRHT are only given for special cases such as small radii \cite{Joulain2005, Ben-Abdallah2011, Dong2017a, Nikbakht2018}, large separation gaps \cite{Kruger2012}, or small skin depths \cite{Volokitin2001, Sasihithlu2011}. Furthermore, Eq. \ref{eq:Transmissivity_SphereEnv} now allows us to probe sphere-environment NFRHT, using the same scattered field coefficients that are necessary to compute for sphere-sphere interactions. Essentially, a second useful quantity can now be determined for free, as a post-processing step.

\section{Cross-Validation of Theoretical Results} \label{sec:CrossValidationofTheoreticalResults}
In this section, we will compare the results of our work against existing methods in the literature. Equations \ref{eq:Transmissivity_SphereSphere} and \ref{eq:Transmissivity_SphereEnv} were implemented in the Wolfram language of Mathematica \cite{Wolfram2016} and validated against T-DDA and BEM. All code and data displayed from here on are available in the supplemental files to this work.

The Mathematica code is written to take the number, outer radii, optical properties, and configuration of the spheres as inputs. Although the code, as attached, assumes identical homogeneous spheres, it may easily be modified to compute NFRHT between layered spheres whose properties are all unique. The lines of code which would require modification are labeled. Furthermore, the algorithm used to compute vector addition translation coefficients is optimized for spheres of approximately equal radii. Though the algorithm works, in principle, for spheres of any size, the time required grows rapidly for spheres with large size disparity. We direct researchers interested in that situation to Sasihithlu and Narayanaswamy \cite{Sasihithlu2014} for asymptotic approximations which improve computation time.

\subsection{Validation Against Thermal Discrete Dipole Approximation}
Edalatpour et al. \cite{Edalatpour2016a} used T-DDA to simulate the NFRHT between three spheres in a chain.  In the notation of this work, Edalatpour et al. simulated three identical spheres with $\rho=0.8$ \si{\micro\meter} and $D = 100$ \si{\nano\meter}. They set $T_{1} = 300$ K and $T_{2}=T_{3} = 0$ K and simulated values of $Q_{j \rightarrow i}$ for $\lambda = 10$ \si{\micro\meter} and two different values of the relative permittivity: $\varepsilon_{res} = -1.36+1.36i$ and $\varepsilon_{non-res} = 9+0.06i$. $\varepsilon_{res}$ and $\varepsilon_{non-res}$ are values of the dielectric function near and away from resonance, respectively.

T-DDA requires volume discretization of the spheres. Edalatpour et al. discretized the outer spheres (1 and 3) into 27,564 non-uniform subvolumes and the middle sphere into 45,800 non-uniform subvolumes. The smallest subvolumes were concentrated at locations nearest to other spheres, to achieve high resolution in the volumes which absorb most heavily. Although it is not mentioned in Edalatpour et al. \cite{Edalatpour2016a}, the authors have stated in other works that T-DDA computation time can be lengthy \cite{Edalatpour2016}. Time, of course, is the price paid for the ability to simulate any geometry.

A comparison of the results of the T-DDA method and this work (labeled DGF) appears in Tables \ref{tab:ThreeSphereResonant} and \ref{tab:ThreeSphereNonResonant}. This work shows excellent agreement with the results of T-DDA. The maximum percent error of the T-DDA results is 2.82\% compared to our DGF results. Edalatpour et al. did not, however, report results for heat transfer to the environment.

\begin{table}
	\caption{\label{tab:ThreeSphereResonant}Comparison of results from current work (DGF) and the thermal discrete dipole approximation (T-DDA) for a three sphere chain using a resonant value of dielectric function. Values from T-DDA are taken from Edalatpour et al. \cite{Edalatpour2016a}.}
	\begin{center}
		\renewcommand{\arraystretch}{1.15}
		\setlength{\tabcolsep}{0.10cm}
		\begin{tabular}{ccccccccc}
			\hline 
			\multicolumn{1}{c}{} &
			\multicolumn{1}{c}{T-DDA (\si{\nano\watt\per\electronvolt})} & 
			\multicolumn{1}{c}{DGF  (\si{\nano\watt\per\electronvolt})} \\
			\hline
			$Q_{1 \rightarrow 2}$ & 197.2 & 193.8 \\ 
			$Q_{1 \rightarrow 3}$ & 2.41 & 2.48 \\ 
			\hline
		\end{tabular} 
	\end{center}
\end{table}

\begin{table}
	\caption{\label{tab:ThreeSphereNonResonant}Comparison of results from current work (DGF) and the thermal discrete dipole approximation (T-DDA) for a three sphere chain using a non-resonant value of dielectric function. Values from T-DDA are taken from Edalatpour et al. \cite{Edalatpour2016a}.}
	\begin{center}
		\renewcommand{\arraystretch}{1.15}
		\setlength{\tabcolsep}{0.10cm}
		\begin{tabular}{ccccccccc}
			\hline 
			\multicolumn{1}{c}{} &
			\multicolumn{1}{c}{T-DDA  (\si{\pico\watt\per\electronvolt})} & 
			\multicolumn{1}{c}{DGF  (\si{\pico\watt\per\electronvolt})} \\
			\hline
			$Q_{1 \rightarrow 2}$ & 0.425 &  0.430 \\ 
			$Q_{1 \rightarrow 3}$ & 0.0285 & 0.0287 \\ 
			\hline
		\end{tabular} 
	\end{center}
\end{table}

\subsection{Validation Against Boundary Element Method}
To compare results for larger spheres and conductance to the environment, numerical computations were performed using {\sc scuff-em}, a free, open-source software implementation of the boundary-element method \cite{scuffem, Reid2015} which models fluctuating-surface-currents as the source of thermal radiation. Three identical spheres of $\rho=10$ \si{\micro\meter} and $D = 1$ \si{\micro\meter} were simulated. These sizes were chosen specifically to be comparable to the thermal wavelength. The optical properties of the spheres chosen to be those of amorphous silicon dioxide, which were given by the built-in Lorentz oscillator model of dielectric function found in {\sc scuff-em}:
\begin{equation}\label{eq:LorentzOscillatorModel}
\varepsilon(\omega) = \varepsilon_{\infty} + \sum_{k} \left[ \frac{S_{k}}{1-\left( \frac{\omega}{\omega_{0,k}} \right)^{2} - i \Gamma_{k} \left( \frac{\omega}{\omega_{0,k}} \right)} \right]
\end{equation}
where all the parameters of the model are given in Table \ref{tab:DielectricFunction}.

\begin{figure}
\includegraphics[width=3 in]{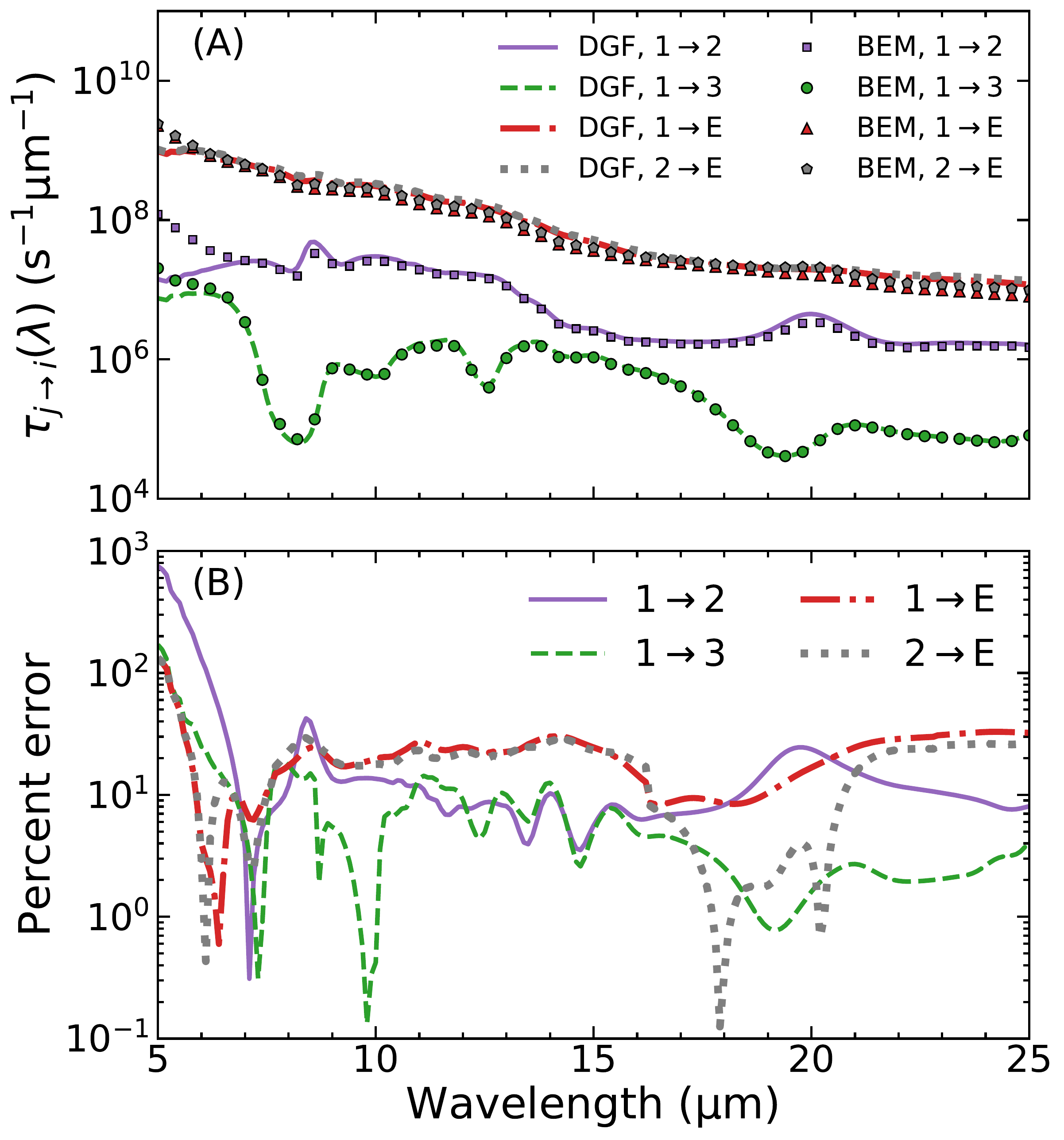}
\caption{\label{fig:ThreeSpheres_SCUFFEM}Cross-validation of DGF and BEM methods. (A) Transmissivity function for energy transfer from source to destination ($j \rightarrow i$) between three identical silicon dioxide spheres with $\rho=10$ \si{\micro\meter} and $D = 1$ \si{\micro\meter}. Values were computed by the DGF (present work) and BEM methods. (B) Magnitude of percent error between two methods in (A).}
\end{figure}

The {\sc scuff-em} software can compute both sphere-sphere and sphere-environment heat transfer. Figure \ref{fig:ThreeSpheres_SCUFFEM}A shows the transmissivity between each pair of spheres and between each sphere and the environment. Due to the symmetry of the three spheres, redundant curves are suppressed. Solid lines are computed using the DGF formulas outlined in this work and markers are from {\sc scuff-em} calculations. For most wavelengths, there is fairly good agreement between the two methods, with approximately 10\% error (see Fig. \ref{fig:ThreeSpheres_SCUFFEM}B). Sphere-sphere errors tend to be lower compared to sphere-environment errors. Additionally, larger errors are apparent for the shortest wavelengths. This may be due to an insufficiently dense mesh. Each sphere's outer surface was meshed into 442 roughly equal triangular areas. Shorter wavelengths should be more sensitive to the deviations from a spherical shape that a mesh introduces and the deviant behavior should be explored further in future work. Fortunately, the $k_{b} (\mathcal{X}/\sinh{\mathcal{X}})^{2}$ term in Eq. \ref{eq:SpectralConductance} is small for short wavelengths, tempering error in total conductance.

\begin{table}
	\caption{\label{tab:DielectricFunction}Lorentz oscillator model parameters for the dielectric function of amorphous silicon dioxide, as given by {\sc scuff-em}. Unlisted is $\varepsilon_{\infty}=2.03843$.}
	\begin{center}
		\renewcommand{\arraystretch}{1.15}
		\setlength{\tabcolsep}{0.10cm}
		\begin{tabular}{ccccccccc}
			\hline 
			\multicolumn{1}{c}{k} &
			\multicolumn{1}{c}{$\hbar \omega_{0,k}$ (eV)} & 
			\multicolumn{1}{c}{$\lambda_{0,k}$ (\si{\micro\meter})} & 
			\multicolumn{1}{c}{$S_{k}$} &
			\multicolumn{1}{c}{$\Gamma_{k}$} \\
			\hline
			1 & 0.05624 & 22.04432 & 0.93752 & 0.09906\\ 
			2 & 0.09952 & 12.45818& 0.05050 & 0.05511 \\
			3 & 0.13355 & 9.28364 & 0.60642 & 0.05246 \\
			\hline
		\end{tabular} 
	\end{center}
\end{table}

\section{Implications for Near-Field Radiative Heat Transfer Experiments} \label{sec:Implications}
Until the recent advances in NFRHT between MEMS devices \cite{Song2016, Cui2017, Fiorino2018}, experiments measuring NFRHT in sub-micron gaps were performed in the microsphere-plane configuration, much like the configurations used in Casimir and van der Waals force experiments \cite{Lamoreaux1997, Mohideen1998, Roy1999, Harris2000}. Recent experimental work investigating the Casimir force \cite{Garrett2018} demonstrates that sphere-sphere geometry is also a feasible configuration to investigate NFRHT. To better understand how such a sphere-sphere NFRHT experiment would work, we must first understand how sphere-plane experiments have been performed.

In past sphere-plane NFRHT experiments \cite{Rousseau2009,Shen2009}, experimenters attached the sphere to a bimaterial microcantilever. Bimaterial cantilevers, such as atomic force microscopy cantilevers, are extremely sensitive calorimeters which deflect with any change in temperature. The planar substrate was fixed at a temperature, either passively to the ambient or heated to a temperature above ambient. If the substrate was fixed at ambient temperature, then the sphere was heated using a laser to create a temperature difference between the two objects. Otherwise, the heated substrate supplied the temperature difference.

The sphere was initally located at some distance above the substrate, typically between 2.5 \si{\micro\meter} and 10 \si{\micro\meter}. The separation between the sphere and the substrate was then decreased until contact was made. Due to surface imperfections and the resolution of the system controlling the separation distance, the minimum distance achievable was approximately 30 \si{\nano\meter}.

As the substrate approaches the sphere, the temperature of the sphere changes which results in deflection of the cantilever; angular deflection is the measurable experimental parameter. Because the initial orientation of a cantilever is large determined by events preceding the experiment (such as fabrication), only the change in the orientation of the cantilever is meaningful and the experiment is sensitive only to changes in total conductance from its value at the initial separation distance to its value at its current separation distance. The angular deflection is a result of all the pathways heat has to flow into and out of the sphere, and it must be related to the change in conductance between the two objects by using a thermal model.

The thermal model used to infer sphere-plane conductance must account for all means of heat transfer occurring. Critically, it must reflect the fact that the sphere-plane system is really a sphere-plane-environment system with exchanges of thermal energy between both objects and their environment. Prior works employing sphere-plane geometries have used a variety of approaches to account for heat transfer to the environment. For example, some works have used Mie theory to compute sphere-environment conductances \cite{Narayanaswamy2008a}, assumed a constant but unnamed value \cite{Shen2009}, ignored environmental heat transfer effects all together \cite{Rousseau2009, Guha2012}, or not reported their treatment of far-field radiation at all \cite{Kim2015}. Any improper treatment of the sphere-environment conductance will introduce a systematic error.

\begin{figure}
\centering
\includegraphics[width=3 in]{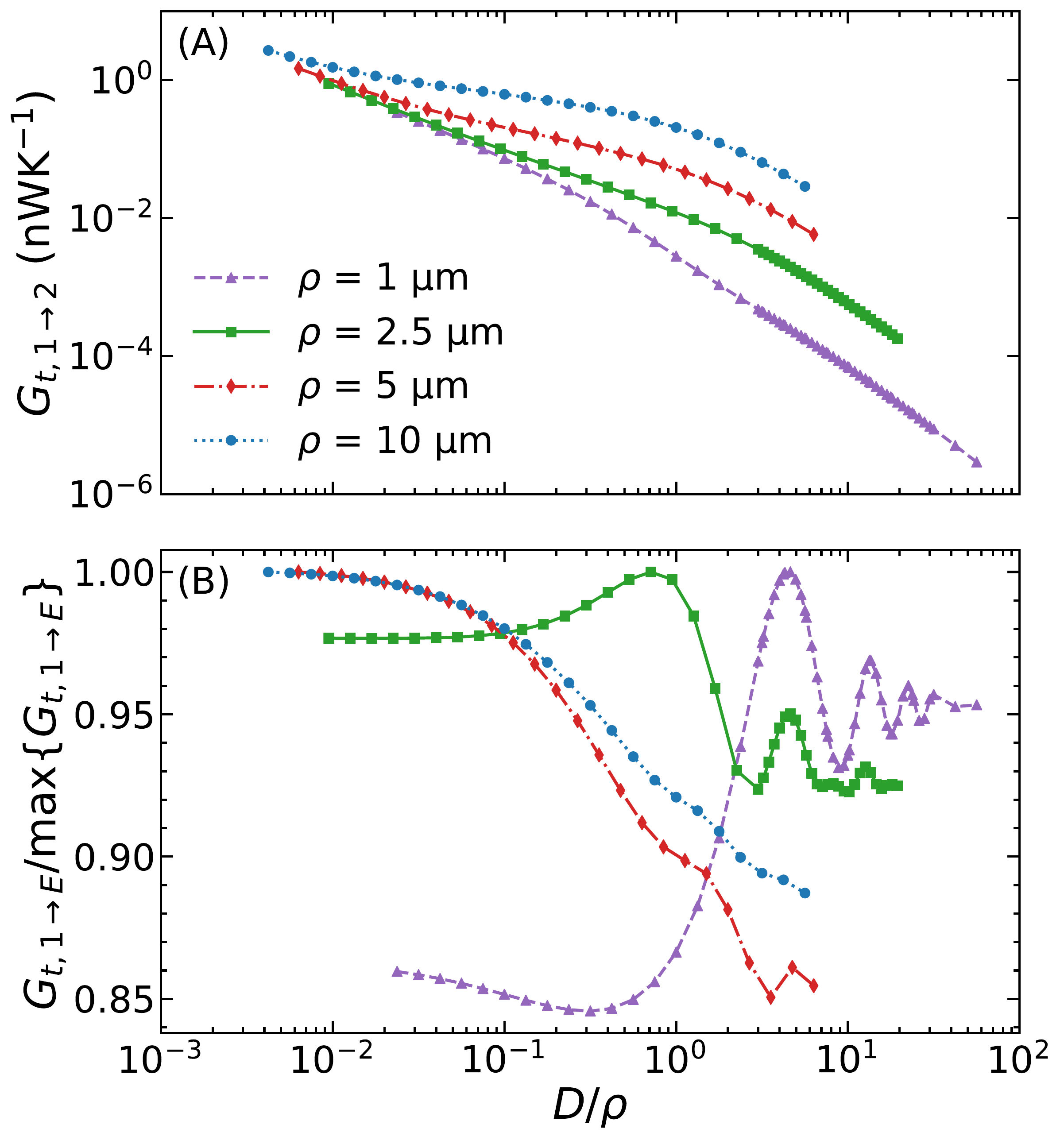}
\caption{\label{fig:DistanceDependence} Total conductance in a system of two identical silicon dioxide spheres with outer radii of 10 \si{\micro\meter}. The legend in (A) is common to both subfigures. (A) Sphere-sphere conductances determined by DGF method. (B) Sphere-environment conductances (normalized by the maximum value of each curve) predicted by the DGF method. See Table \ref{tab:ErrorInSphereSphere} for values of $\mathrm{max}\{G_{t,1 \rightarrow E}\}$.}
\end{figure}

To investigate the magnitude of that error in a potential sphere-sphere experiment, we simulate the NFRHT for two identical silicon dioxide spheres with outer radii of 1, 2.5, 5, and 10 \si{\micro\meter}. Figure \ref{fig:DistanceDependence}A shows sphere-sphere conductances and Fig. \ref{fig:DistanceDependence}B shows normalized sphere-environment conductances. The legend in Fig. \ref{fig:DistanceDependence}A is common to the entire figure. As expected, the sphere-sphere conductances determined by the DGF method show a super-Planckian monotonic decrease in the near-field.

More interestingly, it is readily apparent from Fig. \ref{fig:DistanceDependence}B that the character of the sphere-environment conductances has a strong size dependence. The smallest spheres show strong signs of diffraction for large separation distances, reminiscent of recent work involving point particles (see Fig. 8 by Asheichyk et al. \cite{Asheichyk2017}). The smallest sphere ($\rho$ = 1 \si{\micro\meter}) actually shows a decrease in sphere-environment conductance with decreasing distance. The next smallest sphere ($\rho$ = 2.5 \si{\micro\meter}) shows an eventual increase over its far-field value, though it has a global maximum at an intermediate gap. The largest spheres ($\rho$ = 5 \si{\micro\meter} and 10 \si{\micro\meter}) show near-monotonic decreases and a maximum value at the smallest gap simulated. This demonstrates that the intense electric and magnetic fields between objects which contribute to NFRHT can dampen or enhance far-field emission, with a seeming size dependence. The one common trend is that all sphere-environment conductances eventually level off for sufficiently small $D/\rho$.

The fact that the curves level off is key to taking a valid measurement of sphere-sphere conductance using a cantilever. The angular deflection of a cantilever, $\Delta\theta$, is proportional to the change in total conductances into the sphere attached to the cantilever. Mathematically, that is
\begin{equation} \label{eq:ExperimentalBalance}
\begin{split}
\Delta\theta &\propto G_{t,1\rightarrow 2}(D) - G_{t,1\rightarrow 2}(D_{0}) \\ & \quad + G_{t,1\rightarrow E}(D) - G_{t,1\rightarrow E}(D_{0})
\end{split}
\end{equation}
where $D_{0}$ is the separation distance at which the experiment begins.

Sphere-sphere conductance increases significantly and rapidly in the extreme near-field, such that $G_{t,1\rightarrow 2}(D) \gg G_{t,1\rightarrow 2}(D_{0})$ for most values of $D$ and $D_{0}$. By starting an experiment at a sufficiently small initial separation gap, where the curves in Fig. \ref{fig:DistanceDependence}B have leveled off, $G_{t,1\rightarrow E}(D) - G_{t,1\rightarrow E}(D_{0}) \approx 0$ and an isolation of the sphere-sphere conductance can be achieved.

The largest values of $D_{0}$ for which we can neglect the distance dependence of $G_{t, 1 \rightarrow E}$ and introduce a relative error of no more than approximately 5\% are given in Table \ref{tab:ErrorInSphereSphere}. Although we do not propose a universal criterion at this time, the current results suggests that smaller spheres allow for a larger range of separation distances to be probed. This is contrary to the larger spheres used in past sphere-plane experiments \cite{Rousseau2009,Shen2009}. It can be seen from Table \ref{tab:ErrorInSphereSphere} that, by reducing the radius by one order of magnitude (from 10 \si{\micro\meter} to 1 \si{\micro\meter}), $\mathrm{max}\{ G_{t,1 \rightarrow E} \}$ can be reduced by 2 orders of magnitude. This allows $G_{t, 1 \rightarrow 2}$ to dominate Eq. \ref{eq:ExperimentalBalance} for appropriately chosen sphere sizes. Furthermore, since metallic spheres with coatings of polar materials have been shown to achieve conductances comparable to purely polar spheres in the extreme near-field but lower metal-like conductances at larger gaps (see Fig. 3 by Czapla and Narayanaswamy \cite{Czapla2017}), layered spheres may prove better at suppressing $G_{t,1\rightarrow 2}(D_{0})$ in Eq. \ref{eq:ExperimentalBalance}. All together, these evidence indicate sphere-sphere geometries have great potential as an experimental platform for measuring NFRHT. 

\begin{table}
	\caption{\label{tab:ErrorInSphereSphere}Initial values of $D$ which will ensure a relative error of approximately 5\% or less in a hypothetical sphere-sphere experiment which neglects to quantify the distance dependence of sphere-environment conductance. Also listed are the maximum values of sphere-environment conductance which are used to normalize Fig. \ref{fig:DistanceDependence}B.}
	\begin{center}
		\renewcommand{\arraystretch}{1.15}
		\setlength{\tabcolsep}{0.10cm}
		\begin{tabular}{ccccccccc}
			\hline 
			\multicolumn{1}{c}{$\rho$ (\si{\micro\meter})} &
			\multicolumn{1}{c}{$D_{0}$ (\si{\micro\meter})} &
			\multicolumn{1}{c}{$\mathrm{max}{\{G_{t,1 \rightarrow E}}\}$ (\si{\nano\watt\per\kelvin})} \\
			\hline
			1 & 1.00 & 0.02978 \\ 
			2.5 & 1.00 & 0.3175 \\
			5 & 0.50 & 1.723 \\
			10 & 0.60 & 7.196 \\
			\hline
		\end{tabular} 
	\end{center}
\end{table}

\section{Conclusions}
We theoretically and numerically investigated the NFRHT between pairs of spheres in a linear chain and between individual spheres and their environment using a DGF formalism. The formulas we derived are numerically exact for spheres of any size, spacing, optical properties, and number of spherically symmetric layers. We cross-validated our work against meshed approaches to NFRHT, namely T-DDA and BEM, and demonstrated excellent agreement.

Our work shows a strong distance dependence for the energy radiated by a sphere to its environment in the near-field regime, which levels off in the extreme near-field. Our numerical results should serve as a warning to scientists when analyzing their experimental results and will hopefully spur investigation into more sophisticated thermal models for common NFRHT experimental configurations. Additionally, once implemented, these thermal models can potentially serve as a characterization tool for other factors that impact NFRHT, such as separation distance, variation in optical properties, surface roughness, or surface contamination.

\section*{Acknowledgment}
This work was supported partially by NSF IGERT (DGE-1069240).

\appendix
\section{Solution to Linear System} \label{ap:SolutionToLinearSystem}
At first glance, the coupled system of equations given by Eq. \ref{eq:ScatteredFieldEquations} is not simple to solve for the scattered field coefficients, $V_{l, \nu, m}^{X,Y,i,j}$. To clarify the necessary approach, we start by unpacking Eq. \ref{eq:ScatteredFieldEquations} into all the equations it provides. We fix values of $m$, $i$, and $j$ and generate equations from all possible combinations of $X=M$ or $N$, $Y = M$ or $N$. We obtain
\begin{equation*}
\begin{split}
&  \sum_{p=1}^{N_{s}} \sum_{n = \widetilde{m}}^{\infty}
\left[ \begin{array}{r}
V_{l,n,m}^{M,M,p,j} R_{n}^{(M)}(\rho_{p}) C_{n,\nu,m}^{M,M,i,p}
\\
+ V_{l,n,m}^{N,M,p,j} R_{n}^{(N)}(\rho_{p}) C_{n,\nu,m}^{M,N,i,p}
\end{array} \right]
\\
& \enskip = V_{l, \nu, m}^{M,M,i,j} - C_{l, \nu, m}^{M,M,i,j}
\\
&  \sum_{p=1}^{N_{s}} \sum_{n = \widetilde{m}}^{\infty}
\left[ \begin{array}{r}
V_{l,n,m}^{M,M,p,j} R_{n}^{(M)}(\rho_{p}) C_{n,\nu,m}^{N,M,i,p}
\\
+ V_{l,n,m}^{N,M,p,j} R_{n}^{(N)}(\rho_{p}) C_{n,\nu,m}^{N,N,i,p}
\end{array} \right]
\\
& \enskip = V_{l, \nu, m}^{N,M,i,j} - C_{l, \nu, m}^{N,M,i,j}
\\
&  \sum_{p=1}^{N_{s}} \sum_{n = \widetilde{m}}^{\infty}
\left[ \begin{array}{r}
V_{l,n,m}^{M,N,p,j} R_{n}^{(M)}(\rho_{p}) C_{n,\nu,m}^{M,M,i,p}
\\
+ V_{l,n,m}^{N,N,p,j} R_{n}^{(N)}(\rho_{p}) C_{n,\nu,m}^{M,N,i,p}
\end{array} \right]
\\
& \enskip = V_{l, \nu, m}^{M,N,i,j} - C_{l, \nu, m}^{M,N,i,j}
\\
&  \sum_{p=1}^{N_{s}} \sum_{n = \widetilde{m}}^{\infty}
\left[ \begin{array}{r}
V_{l,n,m}^{M,N,p,j} R_{n}^{(M)}(\rho_{p}) C_{n,\nu,m}^{N,M,i,p}
\\
+ V_{l,n,m}^{N,N,p,j} R_{n}^{(N)}(\rho_{p}) C_{n,\nu,m}^{N,N,i,p}
\end{array} \right]
\\
& \enskip = V_{l, \nu, m}^{N,N,i,j} - C_{l, \nu, m}^{N,N,i,j}
\end{split}
\end{equation*}

In order to solve the linear system numerically, we must truncate the infinite sum to a finite number of terms, which we will denote $N_{\mathrm{max}}$. Additionally, we define $N_{\mathrm{terms}} = N_{\mathrm{max}} - \widetilde{m} + 1$. Now, we define four $N_{\mathrm{terms}} \times N_{\mathrm{terms}}$ matrices of scattered field coefficients: $\overline{\overline{\boldsymbol{V}}}_{m}^{M,M,i,j}$, $\overline{\overline{\boldsymbol{V}}}_{m}^{M,N,i,j}$, $\overline{\overline{\boldsymbol{V}}}_{m}^{N,M,i,j}$, and $\overline{\overline{\boldsymbol{V}}}_{m}^{N,N,i,j}$. Each element in the matrices is a value of $V_{l, \nu, m}^{X,Y,i,j}$ with $l \in [\widetilde{m}, N_{\mathrm{max}}]$ and $\nu \in [\widetilde{m}, N_{\mathrm{max}}]$. The $l$ index increases across rows, and $\nu$ increases down columns. For example, for $m=0$ and $N_{\mathrm{max}}=2$
\begin{equation}
\overline{\overline{\boldsymbol{V}}}_{m=0}^{X,Y,i,j} = \left[ \begin{array}{ccc}
V_{l=1, \nu=1, m=0}^{X,Y,i,j} & V_{l=2, \nu=1, m=0}^{X,Y,i,j} \\
V_{l=1, \nu=2, m=0}^{X,Y,i,j} & V_{l=2, \nu=2, m=0}^{X,Y,i,j}
\end{array}\right]
\end{equation}

Next we define a $(2 N_{\mathrm{terms}}) \times (2 N_{\mathrm{terms}})$ block matrix of coefficients for a given pair of spheres
\begin{equation}
\overline{\overline{\boldsymbol{V}}}_{m}^{i,j} = \left[ \begin{array}{c|c}
\overline{\overline{\boldsymbol{V}}}_{m}^{M,M,i,j} & \overline{\overline{\boldsymbol{V}}}_{m}^{M,N,i,j} \\ \hline
\overline{\overline{\boldsymbol{V}}}_{m}^{N,M,i,j} & \overline{\overline{\boldsymbol{V}}}_{m}^{N,N,i,j}
\end{array}\right]
\end{equation}
and an even greater $(2 N_{s} N_{\mathrm{terms}}) \times (2 N_{s} N_{\mathrm{terms}})$ block matrix for all pairs of spheres, $\overline{\overline{\boldsymbol{V}}}_{m}$, where the rows of the block matrix have increasing values of $i$ and the rows have increasing values of $j$. For example, for a two sphere system
\begin{equation}
\overline{\overline{\boldsymbol{V}}}_{m} = \left[ \begin{array}{c|c}
\overline{\overline{\boldsymbol{V}}}_{m}^{i=1,j=1} & \overline{\overline{\boldsymbol{V}}}_{m}^{i=1,j=2} \\ \hline
\overline{\overline{\boldsymbol{V}}}_{m}^{i=2,j=1} & \overline{\overline{\boldsymbol{V}}}_{m}^{i=2,j=2}
\end{array}\right]
\end{equation}

Now that the organization of the scattered field coefficients is clear, we move on to the organization of the Mie coefficients. Define a column vector, $\boldsymbol{R}^{j}_{\nu}$, whose entries are $R^{(M)}_{\nu}(\rho_{j})$ and then $R^{(M)}_{\nu}(\rho_{j})$, appended together. Only terms for which $\nu \in [\widetilde{m}, N_{\mathrm{max}}]$ are included. For example, for $m=0$ and $N_{\mathrm{max}}=2$
\begin{equation}
\boldsymbol{R}^{j}_{m=0} = \left[ \begin{array}{c}
R^{(M)}_{\nu=1}(\rho_{j}) \\
R^{(M)}_{\nu=2}(\rho_{j}) \\ \hline
R^{(N)}_{\nu=1}(\rho_{j}) \\
R^{(N)}_{\nu=2}(\rho_{j})
\end{array}\right]
\end{equation}

Next, define a Mie coefficient matrix
\begin{equation}
\overline{\overline{\boldsymbol{R}}}_{m} =\overline{\overline{\boldsymbol{I}}} \left[ \begin{array}{c}
\boldsymbol{R}^{j=1}_{m} \\ \hline
\boldsymbol{R}^{j=2}_{m} \\ \hline
\vdots \\ \hline
\boldsymbol{R}^{j=N_{\mathrm{s}}}_{m}
\end{array}\right]
\end{equation}
where $\overline{\overline{\boldsymbol{I}}}$ is an identity matrix with dimensions $(2 N_{s} N_{\mathrm{terms}}) \times (2 N_{s} N_{\mathrm{terms}})$.

The vector addition translation coefficients are first organized into $N_{\mathrm{terms}} \times N_{\mathrm{terms}}$ matrices. Each element in the matrices is a value of the translation coefficient with $l \in [\widetilde{m}, N_{\mathrm{max}}]$ and $\nu \in [\widetilde{m}, N_{\mathrm{max}}]$. The $l$ index increases across rows, and $\nu$ increases down columns. For example, given a pair of spheres $i$ and $j$ and for $m=0$ and $N_{\mathrm{max}}=2$
\begin{equation}
\overline{\overline{\boldsymbol{C}}}_{m=0}^{X,Y,i,j} = \left[ \begin{array}{cc}
C_{l=1,\nu=1,m=0}^{X,Y,i,j} & C_{l=2,\nu=1,m=0}^{X,Y,i,j} \\
C_{l=1,\nu=2,m=0}^{X,Y,i,j} & C_{l=2,\nu=2,m=0}^{X,Y,i,j}
\end{array}\right]
\end{equation}

Next we define a $(2 N_{\mathrm{terms}}) \times (2 N_{\mathrm{terms}})$ block matrix of coefficients for a given pair of spheres
\begin{equation}
\overline{\overline{\boldsymbol{C}}}_{m}^{i,j} = \left[ \begin{array}{c|c}
\overline{\overline{\boldsymbol{C}}}_{m}^{M,M,i,j} & \overline{\overline{\boldsymbol{C}}}_{m}^{M,N,i,j} \\ \hline
\overline{\overline{\boldsymbol{C}}}_{m}^{N,M,i,j} & \overline{\overline{\boldsymbol{C}}}_{m}^{N,N,i,j}
\end{array}\right]
\end{equation}
and an even greater $(2 N_{s} N_{\mathrm{terms}}) \times (2 N_{s} N_{\mathrm{terms}})$ block matrix for all pairs of spheres, $\overline{\overline{\boldsymbol{C}}}_{m}$, where the rows of the block matrix have increasing values of $i$ and the rows have increasing values of $j$. For example, for a two sphere system
\begin{equation}
\overline{\overline{\boldsymbol{C}}}_{m} = \left[ \begin{array}{c|c}
\overline{\overline{\boldsymbol{C}}}_{m}^{i=1,j=1} & \overline{\overline{\boldsymbol{C}}}_{m}^{i=1,j=2} \\ \hline
\overline{\overline{\boldsymbol{C}}}_{m}^{i=2,j=1} & \overline{\overline{\boldsymbol{C}}}_{m}^{i=2,j=2}
\end{array}\right]
\end{equation}

It is good to note here that, due to the definition of $C_{l, \nu, m}^{X,Y,i,j}$ in Eq. \ref{eq:TranslationCoefficients}, all the matrices on the main diagonal of $\overline{\overline{\boldsymbol{C}}}_{m}$ are identically zero.

The solution to the  linear system is given by
\begin{equation}\label{eq:LinearSystem_Matrix}
\begin{split}
\overline{\overline{\boldsymbol{V}}}_{m} &= \overline{\overline{\boldsymbol{R}}}_{m}^{-1} \left[ \overline{\overline{\boldsymbol{I}}} - \overline{\overline{\boldsymbol{R}}}_{m} \overline{\overline{\boldsymbol{C}}}_{m} \right]^{-1} \overline{\overline{\boldsymbol{R}}}_{m} \overline{\overline{\boldsymbol{C}}}_{m}
\\
&= \left[ \overline{\overline{\boldsymbol{I}}} - \overline{\overline{\boldsymbol{C}}}_{m} \overline{\overline{\boldsymbol{R}}}_{m} \right]^{-1} \overline{\overline{\boldsymbol{C}}}_{m}
\end{split}
\end{equation}

The second line of Eq. \ref{eq:LinearSystem_Matrix} is the most compact solution for the unknown scattered field coefficients, as defined in this work. The first line is presented in recognition of the fact that other works on light scattering by spheres sometimes define their scattered field coefficients proportional to $\overline{\overline{\boldsymbol{R}}}_{m} \overline{\overline{\boldsymbol{V}}}_{m}$ \cite{Narayanaswamy2008, Mackowski2008}.

\section*{References}
\bibliographystyle{elsarticle-num}

\end{document}